\documentclass[conference]{IEEEtran}

\usepackage{epsfig}

\newcommand{\CK}{\v Cerenkov}

\begin{document}

\title{Particle identification with the AMS-02 RICH detector: search for
dark matter with antideuterons}

\author{\authorblockN{Lu\'isa Arruda, Fernando Bar\~ao,
\underline{Rui Pereira}}
\authorblockA{LIP/IST \\
         Av. Elias Garcia, 14, 1$^{\textnormal{\scriptsize{o}}}$ andar\\
         1000-149 Lisboa, Portugal \\
         e-mail: pereira@lip.pt}}

\maketitle

\begin{abstract}
The Alpha Magnetic Spectrometer (AMS), whose final version AMS-02 is to be
installed on the International Space Station (ISS) for at least 3 years,
is a detector designed to measure charged cosmic ray spectra with energies
up to the TeV region and with high energy photon detection capability up
to a few hundred GeV, using state-of-the art particle identification
techniques. It is equipped with several subsystems, one of which is a
proximity focusing Ring Imaging \CK\ (RICH) detector equipped with a dual
radiator (aerogel+NaF), a lateral conical mirror and a detection plane
made of 680 photomultipliers and light guides, enabling precise
measurements of particle electric charge and velocity ($\Delta \beta /
\beta \sim$ 10${}^{-3}$ and 10${}^{-4}$ for $Z=$~1 and $Z=$~10~$-$~20,
respectively) at kinetic energies of a few GeV/nucleon. Combining velocity
measurements with data on particle rigidity from the AMS-02 Tracker
($\Delta R / R \sim$ 2\% for $R=$~1~$-$~10 GV) it is possible to obtain a
reliable measurement for particle mass. One of the main topics of the
AMS-02 physics program is the search for indirect signatures of dark
matter. Experimental data indicate that dark, non-baryonic matter of
unknown composition is much more abundant than baryonic matter, accounting
for a large fraction of the energy content of the Universe. Apart from
antideuterons produced in cosmic-ray propagation, the annihilation of dark
matter will produce additional antideuteron fluxes. Detailed Monte Carlo
simulations of AMS-02 have been used to evaluate the detector's
performance for mass separation, a key issue for $\bar{D}/\bar{p}$ separation.
Results of these studies are presented.
\end{abstract}

\section{The AMS-02 experiment}

\begin{figure}[htb]

\center


\mbox{\epsfig{file=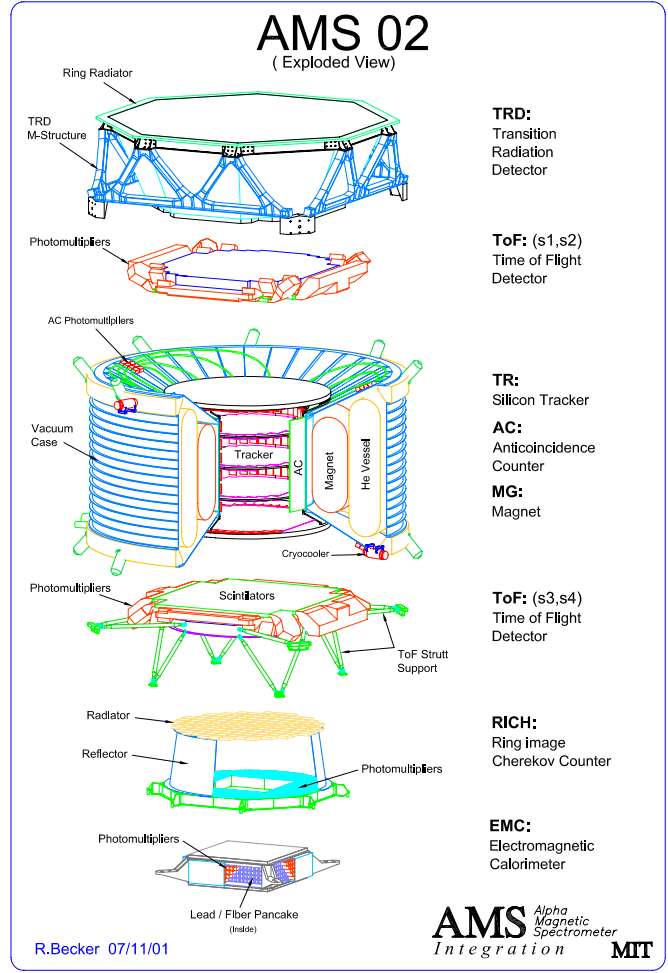,width=0.48\textwidth,clip=}}

\caption{Exploded view of the AMS-02 detector.\label{amsdet}}


\end{figure}

The Alpha Magnetic Spectrometer (AMS)\cite{bib:ams}, whose final version
AMS-02 is to be installed on the International Space Station (ISS) for at
least 3 years, is a detector designed to study the cosmic ray flux by
direct detection of particles above the Earth's atmosphere using
state-of-the-art particle identification techniques. AMS-02 is equipped
with a superconducting magnet cooled by superfluid helium. The
spectrometer is composed of several subdetectors: a Transition Radiation
Detector (TRD), a Time-of-Flight (TOF) detector, a Silicon Tracker,
Anticoincidence Counters (ACC), a Ring Imaging \CK\ (RICH) detector and an
Electromagnetic Calorimeter (ECAL). Fig.~\ref{amsdet} shows a schematic
view of the full AMS-02 detector. A preliminary version of the detector,
AMS-01, was successfully flown aboard the US space shuttle Discovery in
June 1998.

The main goals of the AMS-02 experiment are:

\begin{itemize}

\item A precise measurement of charged cosmic ray spectra in the rigidity
region between \mbox{$\sim$ 0.5 GV} and \mbox{$\sim$ 2 TV}, and the detection
of photons with energies up to a few hundred GeV;

\item A search for heavy antinuclei ($Z \ge$ 2), which if discovered would
signal the existence of cosmological antimatter;

\item A search for dark matter constituents by examining possible
signatures of their presence in the cosmic ray spectrum.

\end{itemize}

The long exposure time and large acceptance (0.5 m${}^2$sr) of
AMS-02 will enable it to collect an unprecedented statistics of more than
$10^{10}$ nuclei.

\begin{figure}[htb]

\center


\mbox{\epsfig{file=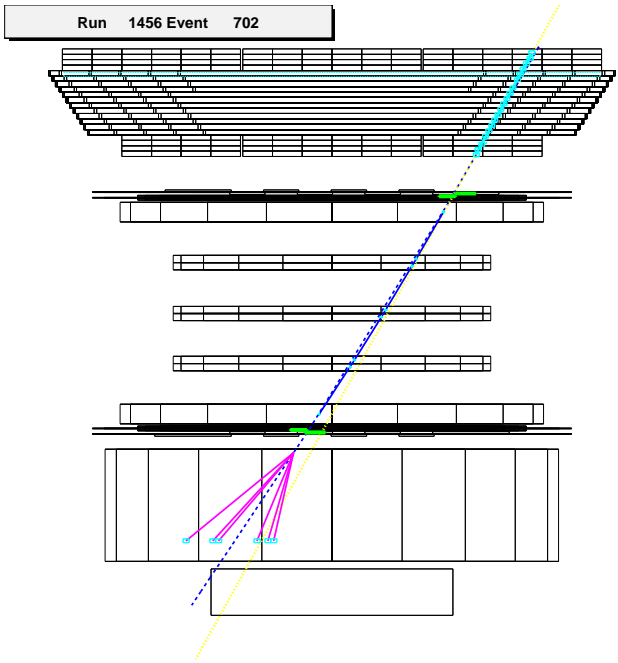,width=0.48\textwidth,clip=}}

\vspace{-0.3cm}

\caption{A simulated proton event as seen in the AMS-02 display.\label{amsdisplay1}}

\end{figure}

\section{The AMS RICH detector}

One of the subdetectors in AMS-02 is a proximity focusing Ring Imaging
\CK\ (RICH) detector. It is composed of a dual radiator with silica
aerogel ($n=$~1.050) and sodium fluoride ($n=$~1.334), a high reflectivity
lateral conical mirror and a detection matrix with 680 photomultipliers
coupled to light guides.

\begin{figure}[htb]

\center

\vspace{-0.5cm}

\mbox{\epsfig{file=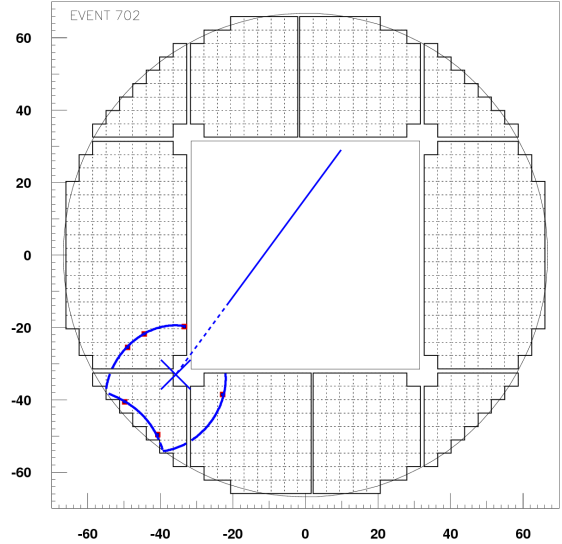,width=0.48\textwidth,clip=}}

\vspace{-0.3cm}

\caption{The same event of Fig.~\ref{amsdisplay1} as seen in the
RICH display developed at LIP.\label{amsdisplay2}}

\end{figure}

The RICH detector will provide a very accurate velocity measurement (in
aerogel, $\Delta \beta / \beta \sim$ 10${}^{-3}$ and 10${}^{-4}$ for
$Z=$~1 and $Z=$~10~$-$~20, respectively) and charge identification of
nuclei up to iron ($Z=$~26).

RICH data, combined with information on particle rigidity from the AMS
Silicon Tracker, enable the reconstruction of particle mass. A typical
RICH event is shown in Figs.~\ref{amsdisplay1} and \ref{amsdisplay2},
where the latter gives a detailed view of the readout matrix. The accuracy
of the RICH velocity measurement is essential due to the growth of
relative errors when $v \to c$:

\begin{displaymath}
\frac{\Delta m}{m} = \frac{\Delta p}{p} \oplus \gamma^2 \frac{\Delta
\beta}{\beta}
\end{displaymath}

The assembly of the AMS RICH detector is currently underway at CIEMAT in
Madrid. The integration of the RICH and the other subdetectors of AMS-02
will take place at CERN in 2007.

The analysis of RICH data involves the identification of the \CK\ ring in
a hit pattern which usually includes several scattered noise hits and an
eventual strong spot in the region where
the charged particle crosses the detection plane. Two independent algorithms
for velocity and charge reconstruction have been developed in the AMS
collaboration for the analysis of RICH events: a geometrical method based
on a hit-by-hit reconstruction\cite{bib:elisa}, and a method using all the
hits with the maximization of a likelihood function\cite{bib:rich2003}).

A prototype of the RICH detector, consisting of 96 photomultiplier units,
was tested both with cosmic ray particles and with beam ions at the CERN SPS
in 2002 and 2003. A piece of the conical reflector was included in the
beam test setup\cite{bib:prototype}. The algorithms for velocity and charge
reconstruction were successfully applied to data from these prototype
tests\cite{bib:luisa}.

\section{Dark matter and the antideuteron signal}

Dark matter has been the subject of astrophysical research since the
first half of the 20${}^{\textnormal{th}}$ century. Measurements of
galactic rotation curves and of relative velocities of objects in galaxy
clusters have shown that the total mass of galaxies and clusters is much
higher than the mass of what is directly observed as luminous
matter\cite{bib:mosca}.

In recent years major progress has been made on this subject.
The most recent cosmological data from WMAP\cite{bib:wmap} indicate that
baryons account for only a small fraction of the total matter density of
the Universe ($\Omega_b \approx$~0.04, $\Omega_m \approx$~0.24). The
remaining mass should correspond to particles that have not been observed
yet.

\begin{figure}[htb]

\center


\mbox{\epsfig{file=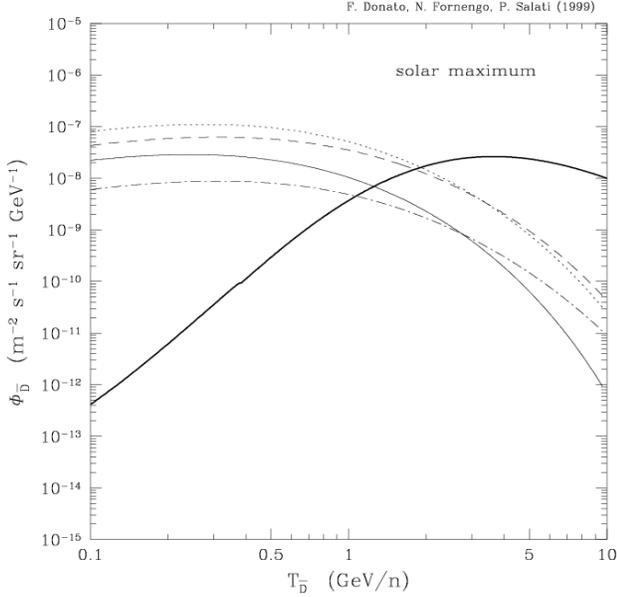,width=0.48\textwidth,clip=}}

\caption{Comparison of expected antideuteron fluxes from secondary
processes and neutralino annihilation: solar maximum.
The four cases shown for $\chi$ annihilation correspond
to different sets of parameters in the context of the Minimal
Supersymmetric extension of the Standard Model (MSSM)
(from Ref.~\protect \cite{bib:antideut}) \label{antidfluxmax}}

\vspace{-0.4cm}

\end{figure}

\begin{figure}[htb]

\center


\mbox{\epsfig{file=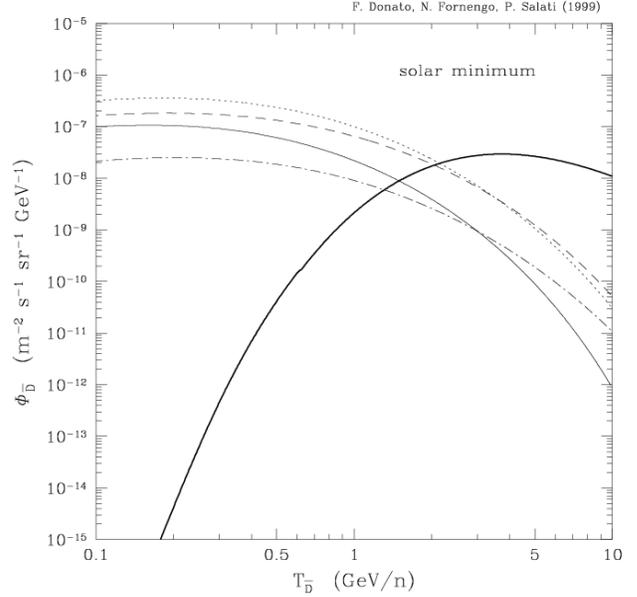,width=0.48\textwidth,clip=}}

\caption{Same as fig. \ref{antidfluxmax}, for solar minimum.
(from Ref.~\protect \cite{bib:antideut}) \label{antidfluxmin}}


\end{figure}

The neutralino ($\chi$), a heavy, neutral, stable particle predicted by
supersymmetric models, is a favourite dark matter candidate.
If supersymmetry exists, the annihilation of neutralino pairs is expected
to produce a significant effect on certain components of the cosmic ray
spectrum (namely $\gamma$, $e^+$, $\bar{p}$, $\bar{D}$). In particular,
the low energy antideuteron flux resulting from neutralino annihilation is
expected to be orders of magnitude higher than the secondary flux due to
other interactions\cite{bib:antideut}, as shown in Figs.~\ref{antidfluxmax}
and \ref{antidfluxmin}.

\section{Particle identification: the deuteron case}

To evaluate the capabilities of AMS-02 for mass separation of
antideuterons from other particles with the same charge, studies have been
performed using the similar case of deuteron vs. proton separation. In the
past, studies on the separation of helium ($Z=$~2) and beryllium ($Z=$~4)
isotopes have also been performed using a standalone simulation of the
RICH detector\cite{bib:thesisluisa}. In the present case the
large difference between proton and deuteron abundances (D/p~1\%) increases
the importance of a very effective mass separation to isolate the deuteron
signal from a large background of proton events.

In the study of D/p separation a full-scale simulation of the AMS detector
was used. Particles were simulated as coming from the top plane of a cube,
corresponding to an acceptance of 47.78 m${}^2$sr. Three data samples were
chosen. Table \ref{simstat} shows the momentum ranges and number of events
simulated in each sample.

\begin{table}[htb]

\vspace{-0.3cm}

\begin{center}

\caption{Samples used in the $D/p$ separation studies}
{
\begin{tabular}{|c|c|c|}

\hline \textbf{Sample} & \textbf{Momentum range} & \textbf{No. events}
\\ \hline $p$ (low momentum) & 0.5~$-$~10 GeV/c & 3.1~$\times$~10${}^8$
\\ \hline $p$ (high momentum) & 10~$-$~200 GeV/c & 1.3~$\times$~10${}^8$
\\ \hline $D$ & 0.5~$-$~20 GeV/c & 5.6~$\times$~10${}^7$

\\ \hline \multicolumn{3}{c}{}

\end{tabular}

\label{simstat}
}

\end{center}

\vspace{-0.5cm}

\end{table}

\begin{figure}[htb]

\center


\mbox{\epsfig{file=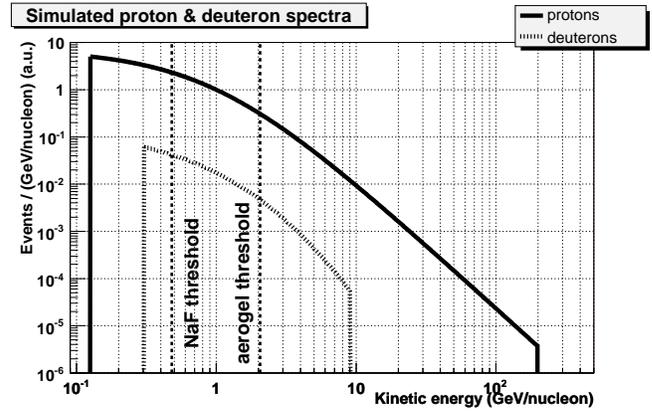,width=0.48\textwidth,clip=}}

\caption{Simulated proton and deuteron spectra used in this work.
\label{simspec}}

\vspace{-0.4cm}

\end{figure}

For each sample, $\frac{dN}{d(ln \, p)} =$~constant. Variable
weights were assigned to events in order to compensate for the statistics
in each sample and to reproduce a realistic spectrum (Fig.~\ref{simspec}):

\begin{itemize}

\item The simulated proton spectrum followed \mbox{$dN/dE \propto E^{-2.7}$};

\item The simulated deuteron spectrum was calculated combining the proton
spectrum above with D/p ratios taken from Ref.~\cite{bib:seo:h}.

\end{itemize}

In each event a set of preliminary data selection cuts using readings from
different subdetectors of AMS-02 was applied to reduce the fraction of
events with a bad reconstruction. Only downgoing events ($\beta > 0$)
were accepted. In addition, events were accepted if the following conditions
were satisfied:

\begin{itemize}

\item Only one particle was detected in the event;

\item A particle track was reconstructed by the Silicon Tracker;

\item No clusters were found in the Anti-Coincidence Counters;

\item Clusters from at least 3 TOF planes (out of 4) were used for event
reconstruction;

\item At most one additional cluster was allowed in the TOF; 

\item At least 6 Tracker layers (out of 8) were used in the track
reconstruction;

\item Compatibility was required for the rigidity measurements obtained from
two different algorithms, with \mbox{$\Delta R/R < 3\%$};

\item Compatibility was also required for the rigidity measurements obtained
from each half of the Tracker (upper and lower), with \mbox{$\Delta R/R < 50\%$};

\item The particle's impact point on the RICH radiator was less than 58 cm
from the centre (i. e. more than 2 cm from the mirror);

\item At most one track was present in the TRD;

\item The TOF and Tracker charge reconstructions were compatible.

\end{itemize}

Among the events that triggered the detector, a fraction corresponding to
\mbox{$\sim$~15-20\%} of proton events and \mbox{$\sim$~10-15\%}
of deuteron events in the relevant region of kinetic energy (few GeV/nucleon)
passed this set of preliminary cuts, corresponding to an acceptance of
$\sim$~0.3 m${}^2$sr for protons and $\sim$~0.2 m${}^2$sr for deuterons.

The reconstruction of particle masses was then performed for events having
a signal in the RICH detector. The extremely accurate velocity measurement
provided by the RICH ($\Delta \beta / \beta \sim$ 10${}^{-3}$ in the case
of protons and deuterons) is crucial to reduce the background level. A series
of event selection cuts were introduced, based on data provided by the RICH
and the results of the two reconstruction algorithms:

\begin{figure}[htb]

\center


\mbox{\epsfig{file=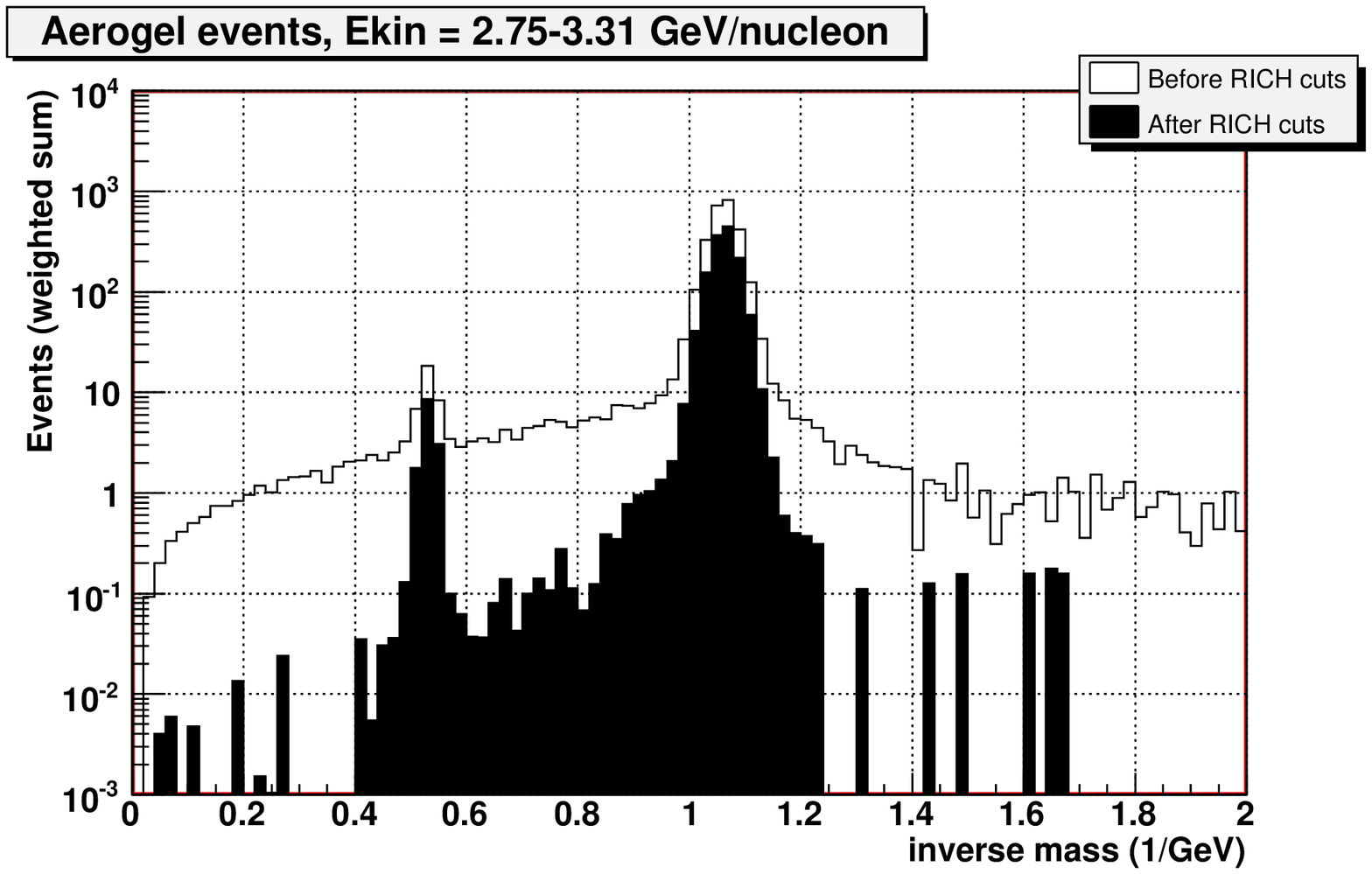,width=0.48\textwidth,clip=}}

\mbox{\epsfig{file=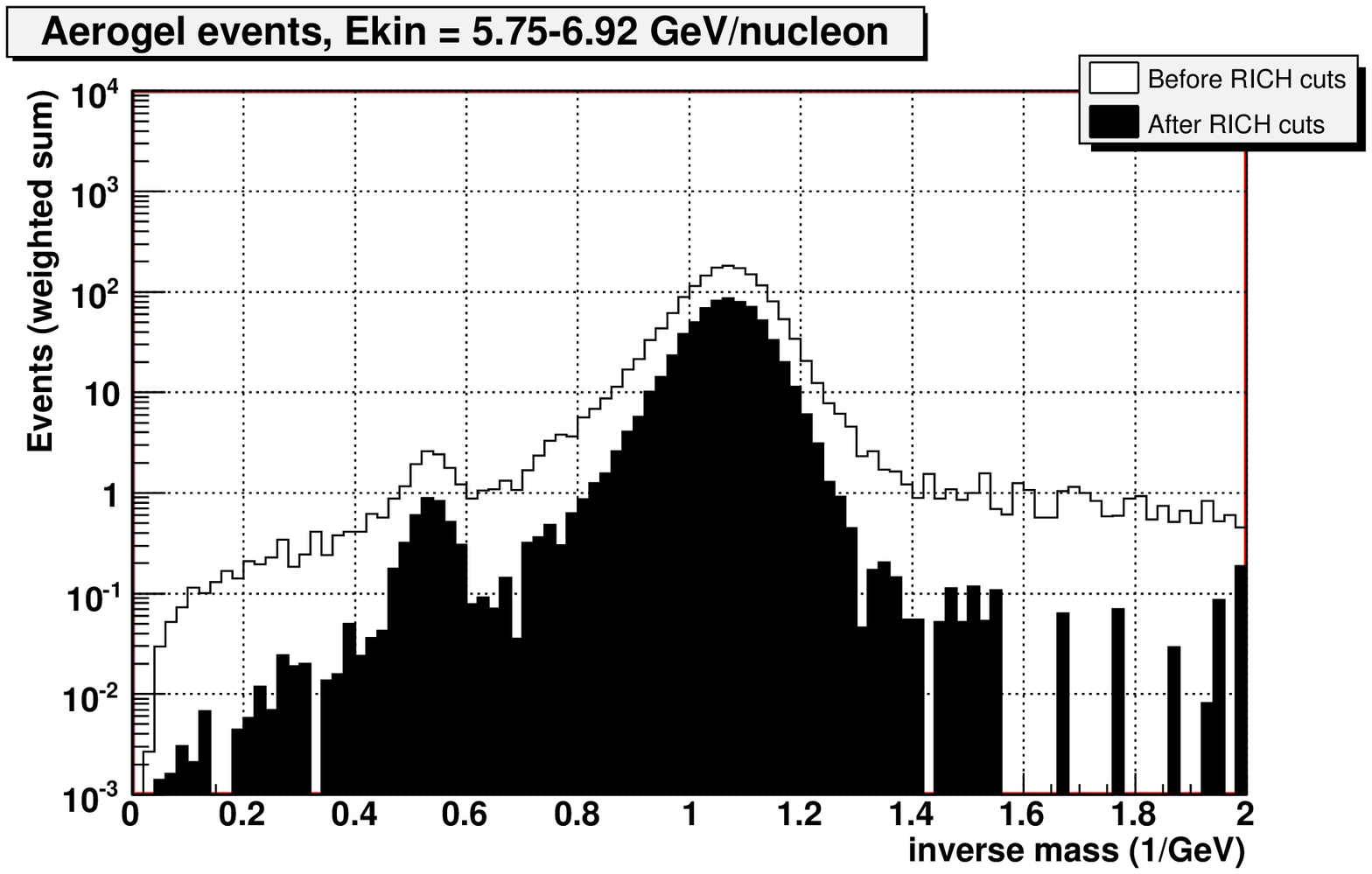,width=0.48\textwidth,clip=}}


\caption{Examples of inverse mass distribution in aerogel events for two
energy regions.
\label{imassagl}}

\end{figure}

\begin{itemize}

\item A \CK\ ring was reconstructed using each method, and at least 3 hits
were used in both cases;

\item The number of ring hits was not higher than 10 in NaF events, and not
higher than 15 in aerogel events;

\item A Kolmogorov test to the uniformity of the hits azimuthal distribution
in the ring gave a result of at least 0.2 in the case of NaF events, and 0.03
in the case of aerogel events;

\item Compatibility was required for the velocity measurements from the TOF
and RICH detectors, with \mbox{$\Delta \beta/\beta < 10\%$};

\item Compatibility was also required for the velocity measurements obtained
from the two RICH reconstruction methods, with \mbox{$\Delta \beta/\beta < 0.3\%$}
for NaF events, and \mbox{$\Delta \beta/\beta < 0.07\%$} for aerogel events;

\item The reconstructed, rounded electric charge obtained from the geometrical
method was 1 or 2;

\item The reconstructed, non-rounded electric charge obtained from the
likelihood method was between 0.5 and 1.5 in NaF events, and between 0.6
and 1.4 in aerogel events;

\item The ring acceptance (visible fraction), as estimated by
the likelihood method, was at least 20\% in NaF events, and at least 40\% in
aerogel events;

\item The number of noisy hits not associated to the crossing of the charged
particle (i. e., hits that were far from the reconstructed ring and far from
the estimated crossing point of the charged particle in the detection matrix)
was not higher than 2 in NaF events, and not higher than 4 in aerogel events.

\end{itemize}

\section{Analysis results}

\begin{figure}[htb]

\center


\mbox{\epsfig{file=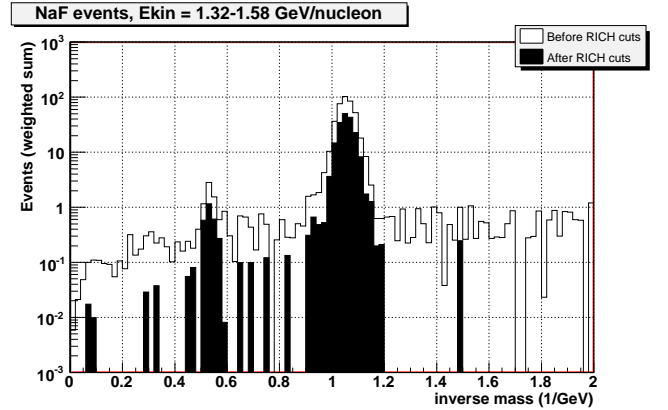,width=0.48\textwidth,clip=}}


\caption{Example of inverse mass distribution in NaF events.
\label{imassnaf}}

\end{figure}

Results show that mass separation of particles with $Z=$~1 is feasible even
if one species is orders of magnitude more abundant than the other. D/p
separation is possible up to $E_{kin} \sim$~8 GeV/nucleon. Some examples
of the mass distributions obtained are shown in Figs.~\ref{imassagl} and
\ref{imassnaf}. Solid lines show the mass distributions before the RICH
cuts were taken into consideration.

\begin{figure}[htb]

\center


\mbox{\epsfig{file=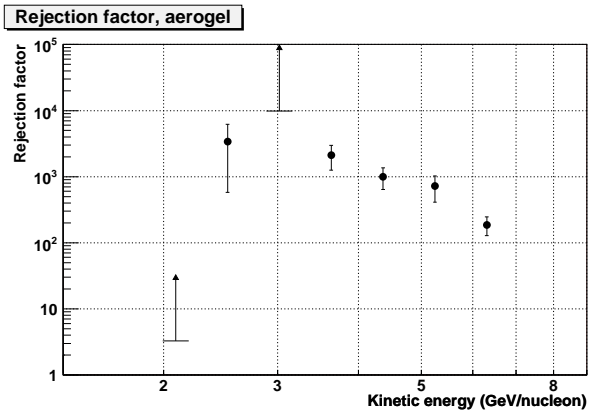,width=0.48\textwidth,clip=}}


\caption{Rejection factor for D/p separation in aerogel events.
\label{rejfac}}

\end{figure}

\begin{figure}[htb]

\center


\mbox{\epsfig{file=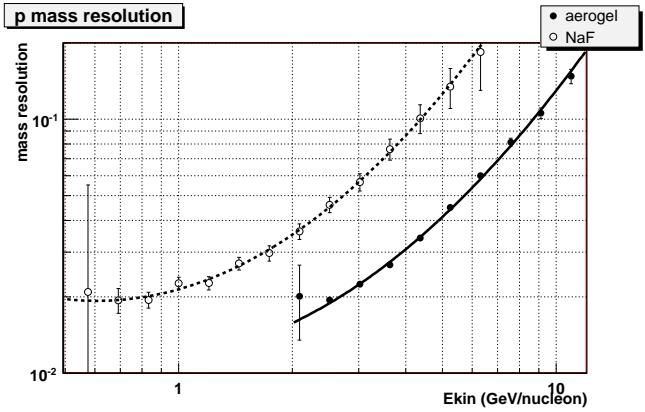,width=0.48\textwidth,clip=}}


\caption{Relative mass resolution for protons: NaF events (open dots)
and aerogel events (filled dots).
\label{massres}}

\vspace{-0.3cm}

\end{figure}

In the optimal region immediately above the aerogel
radiation threshold ($E_{kin} =$~2.1~$-$~4 GeV/nucleon) rejection factors in
the 10${}^3$~$-$~10${}^4$ region were attained (Fig.~\ref{rejfac}). The
best relative mass resolutions for protons (Fig.~\ref{massres}) and deuterons
are $\sim$~2\% for both radiators in the regions above their respective
thresholds.

\begin{figure}[htb]

\center


\mbox{\epsfig{file=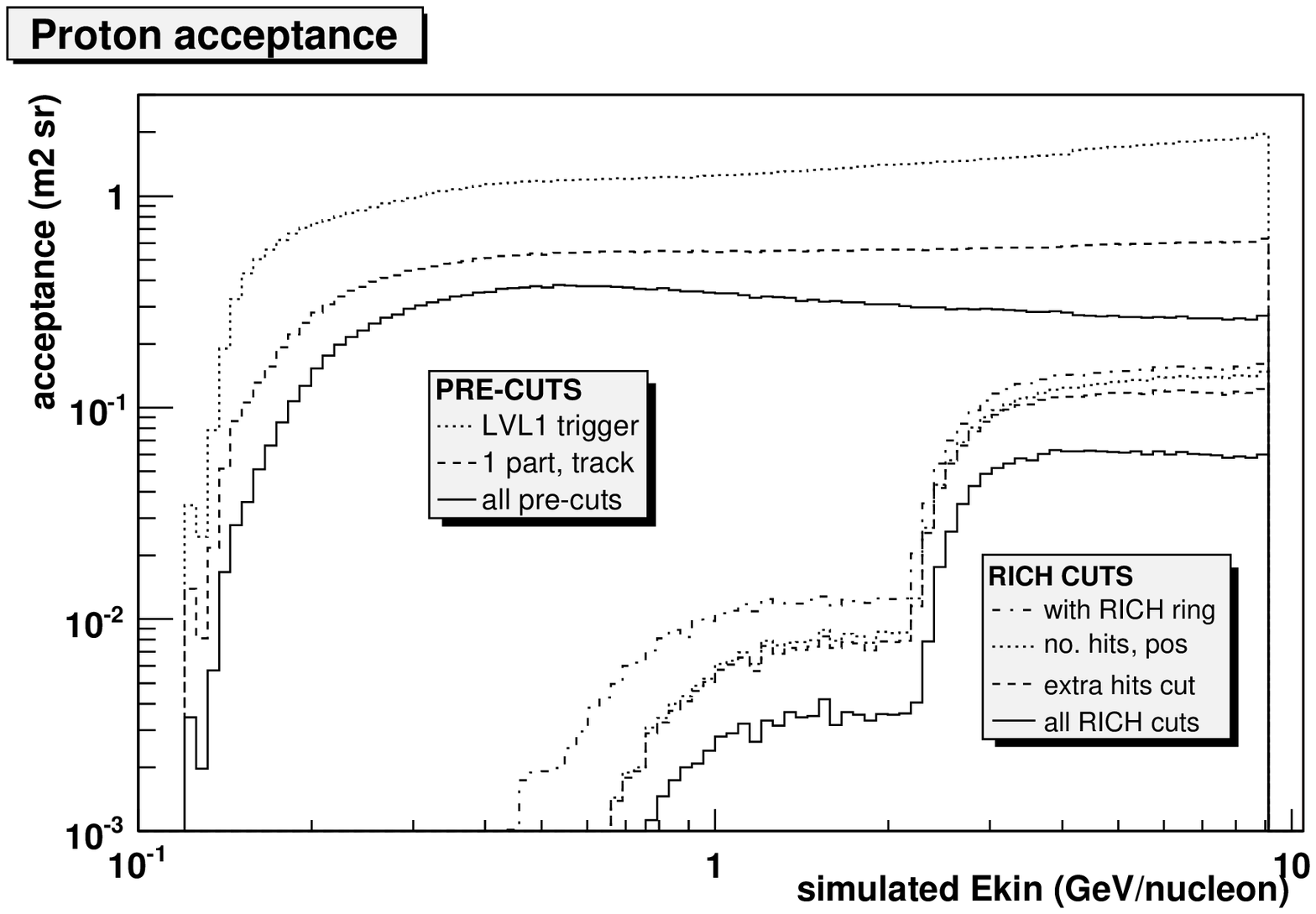,width=0.48\textwidth,clip=}}

\mbox{\epsfig{file=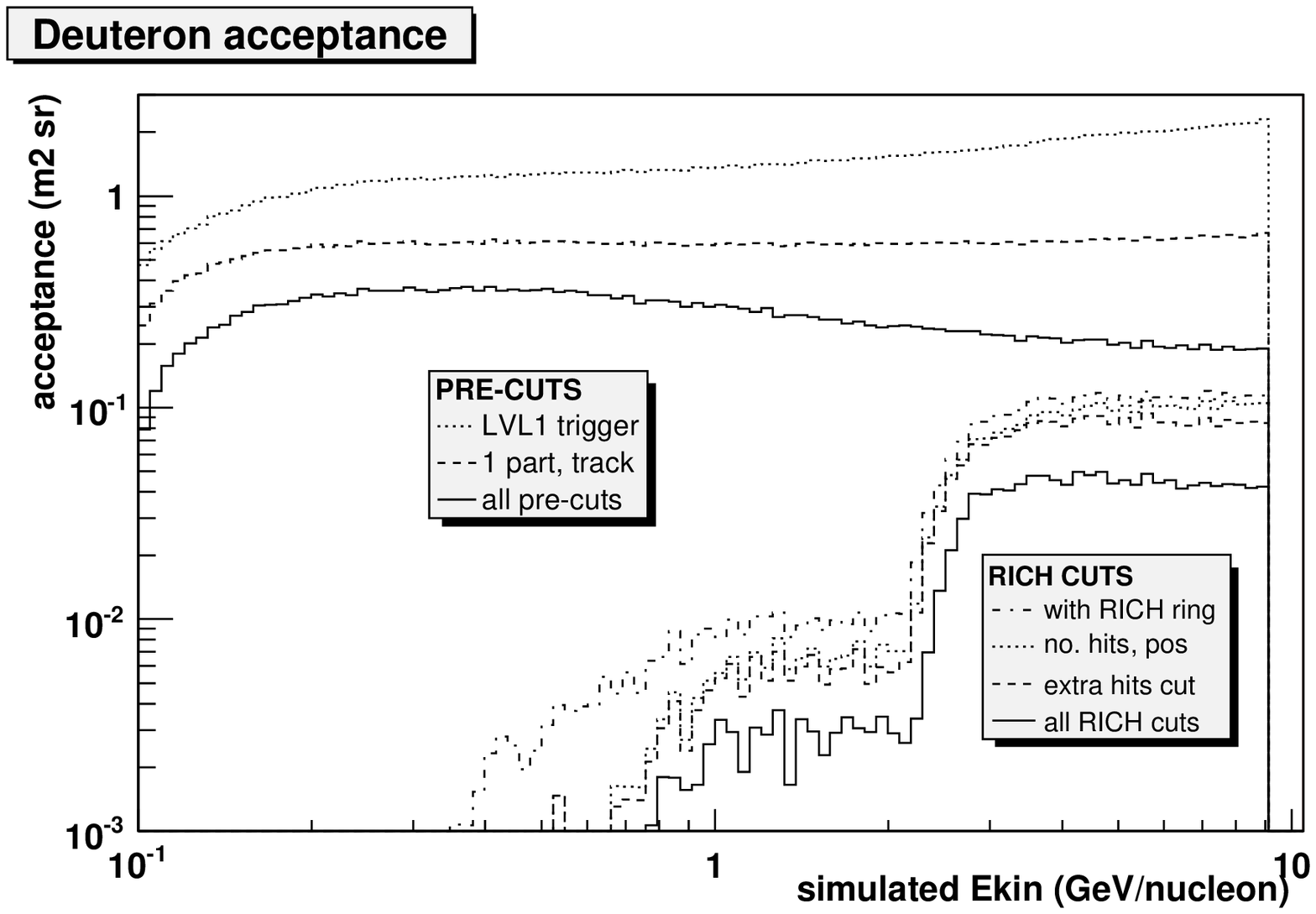,width=0.48\textwidth,clip=}}


\caption{Acceptance for protons (top) and deuterons (bottom) at different
stages of event analysis. Solid lines correspond to acceptances after the
preliminary cuts (third line from top) and after all cuts (lower line).
\label{pdaccep}}

\end{figure}

After all cuts, an acceptance of $\sim$~0.06 m${}^2$sr was
obtained for protons, and $\sim$~0.04 m${}^2$sr for deuterons at
\mbox{$E_{kin} >$~3 GeV/nucleon} (Fig.~\ref{pdaccep}). The increase by a factor
$\sim$~10 in the acceptance above the aerogel threshold reflects the
relative dimensions of the two radiators in the RICH detector.

The main background in the deuteron case comes from non-gaussian tails of
proton events with a bad velocity reconstruction. Errors in rigidity
reconstruction ($\Delta R / R \sim$~2\% in the GeV region) are not
critical for this case.

The specific set of cuts shown here corresponds to an example of a
selection procedure. Other variations are possible. In particular,
rejection factors may be improved by applying stricter cuts, at the
expense of a further acceptance reduction.

\section{Conclusions}

AMS-02 will provide a major improvement on the current knowledge of cosmic
rays. A total statistics of more than 10${}^{10}$ events will be collected
during its operation. Detailed simulations have been performed to evaluate
the detector's particle identification capabilities, in particular those
of the RICH, which might be crucial for the identification of an
antideuteron flux resulting from neutralino annihilation. Simulation
results show that the separation of light isotopes is feasible. Using a
set of simple cuts based on event data, relative mass resolutions of
$\sim$~2 \% and rejection factors up to 10${}^4$ have been attained in D/p
separation at energies of a few GeV/nucleon.

\end{document}